\documentclass{PoS}
\usepackage{wrapfig}

\title{
\begin{picture}(0,0)(0,0)%
   \put(230,75){\makebox(0,0)[l]{\textnormal
{\normalsize OU-HET-764-2012, UTHEP-645
}}}%
\end{picture}%
Pion form factors in the $\epsilon$ regime}

\ShortTitle{Pion form factors in the $\epsilon$ regime}

\author{JLQCD Collaboration: 
        \speaker{H.~Fukaya}$^a$\thanks{E-mail: hfukaya@het.phys.sci.osaka-u.ac.jp},
        S.~Aoki$^{b,c}$,
        S.~Hashimoto$^{d,e}$,
        T.~Kaneko$^{d,e}$,
        H.~Matsufuru$^{d,e}$,
        J.~Noaki$^{d}$,
        T.~Onogi$^a$
        and
        N.~Yamada$^{d,e}$
        \\
        \\
        \\
        \llap{$^a$}
        Department of Physics, Osaka University, 
        Toyonaka, Osaka 560-0043 Japan
        \\
        \llap{$^b$}
        Graduate School of Pure and Applied Sciences, 
        University of Tsukuba, Tsukuba 305-8571, Japan
        \\ 
        \llap{$^c$}
        Center for Computational Sciences, University of Tsukuba, 
        Tsukuba 305-8577, Japan
        \\
        \llap{$^d$}
        High Energy Accelerator Research Organization (KEK),
        Tsukuba 305-0801, Japan 
        \\
        \llap{$^e$}
        School of High Energy Accelerator Science,
        The Graduate University for Advanced Studies (Sokendai),
        Tsukuba 305-0801, Japan
}

\abstract{
We consider how to extract the pion form factors 
in the $\epsilon$ regime.
Using the correlators with non-zero momenta
and taking appropriate ratios of them, we eliminate
the dominant finite volume effect from
the zero-momentum pion mode.
Our preliminary lattice result for the pion charge radius
is consistent with the experiment.
}

\FullConference{The 30 International Symposium on Lattice Field Theory - Lattice 2012,\\
		June 24-29, 2012\\
		Cairns, Australia}

\begin{document}

\section{Introduction}
\label{sec:intro}

JLQCD collaboration has been performing lattice QCD simulations
with dynamical overlap fermions \cite{Neuberger:1997fp}. 
The exact chiral symmetry realized on the lattice by the overlap fermions, 
not only enables us to 
analyze QCD in a theoretically clean way, but also 
makes the hybrid Monte Carlo updates fairly stable 
even in the vicinity of the chiral limit.
In fact, we have succeeded in simulating QCD with 
the pion mass below its physical point 
\cite{Fukaya:2007fb, Fukaya:2007pn, Fukaya:2009fh, Fukaya:2011in}. 

A drawback of the use of the overlap fermion action
is its high numerical cost.
We therefore had to choose a rather small lattice size
$L\sim 1.8$ fm, except for two main simulations
with the pion masses around $290$ MeV and $380$ MeV,
for which we set $L \sim 2.7$ fm.
In particular, our simulation with the lightest pion 
mass $\sim 100$ MeV is inside the so-called $\epsilon$ regime \cite{Gasser:1986vb},
where the correlation length of the pion exceeds the lattice size $L$.

In the $\epsilon$ regime, the finite volume effects
become large, so that the physics
is largely distorted from the infinite volume limit.
Such finite volume effects are mainly due to the pion's zero-momentum mode, 
and can be described well within chiral perturbation theory (ChPT)
with the zero-mode treated non-perturbatively.
We have found that the chiral perturbation formulas
for the Dirac eigenmode density \cite{Damgaard:2008zs}, 
and that for the pseudoscalar
two point functions \cite{Aoki:2011pza}, 
nicely describe our lattice data.
We have extracted the chiral condensate \cite{Fukaya:2009fh}, 
as well as the pion mass and decay constant \cite{Fukaya:2011in}. 

A similar analysis using the chiral expansion
is possible also for the 
three(or more)-point functions \cite{Giusti:2004an}.
It is, however, expected that the contribution from the pion zero-mode 
become more complicated containing non-trivial 
combinations of (modified) Bessel functions, 
as is already seen in the one- and two-point functions.

In this work on the pion form factors,
we propose a different direction,
or a greatly simplified way of analysis 
in the $\epsilon$ regime of QCD.
The key ideas are
\begin{enumerate}
\item To insert non-zero momenta to relevant operators.
\item To take appropriate ratios of them.
\end{enumerate}
Then, we can automatically eliminate the leading ${\cal O}(1)$
finite volume effects.
Namely, we can extract the pion form factors
without relying on non-trivial Bessel functions, 
as in the perturbative regime ($p$ regime) \cite{Aoki:2009qn}.
Of course, this cancellation happens only at the leading order,
and the next-to-leading order terms
may contain non-trivial zero-mode contributions.
But we will show here that the tree-level calculation 
already describe our lattice data for the vector form factor reasonably well,
yielding a consistent value of the pion charge radius with the experiment.

\section{Two-point functions in the $\epsilon$ regime}
\label{sec:2pt}

Before considering the three-point functions,
let us begin with a simpler case with the two-point correlators
to illustrate our new idea.
For simplicity, we consider a two-flavor theory in a finite volume $V=L^3 T$,
with a degenerate quark mass $m$.
The boundary condition is set periodic in every direction.
We denote the chiral condensate by $\Sigma$ and the pion decay constant by $F$.
Note that including the (sea) strange quark 
is not difficult \cite{Bernardoni:2008ei}
and does not change the following results 
at the leading order of ChPT.

The $\epsilon$ expansion of ChPT is expressed by
an exact group integration of the pion's zero-momentum mode,
as well as perturbative non-zero mode fields, 
where their mass is treated as a perturbation.
Namely, the theory is 
a hybrid system of an $SU(2)$ matrix model 
(or a U(2) matrix model when the global topological charge of the gauge fields 
is fixed.), and massless fields.

The pseudoscalar correlator separated by
a four vector $x=(t, x_1,x_2,x_3)$ 
is, thus, given by \cite{Bernardoni:2008ei}
\begin{eqnarray}
\langle P(x)P(0) \rangle &=& X 
+ Y \left(\frac{1}{V}\sum_{p \neq 0} \frac{e^{ipx}}{p^2}\right)+
Z \left(\frac{1}{V}\sum_{p \neq 0} \frac{e^{ipx}}{(p^2)^2}\right)+\cdots ,
\end{eqnarray}
where $X,Y,Z\cdots$ denote (dimensionful) constants, including
non-trivial (Bessel) functions of 
$m \Sigma V$ arising from the pion zero-mode.
The propagator-like forms are contributions from 
the non-zero modes.
Note that the only non-zero momenta are summed.

From the massless boson-like structure, it is not surprising 
to see that the zero-momentum correlator
is a polynomial function of $t$.
This is quite different from 
a conventional hyperbolic cosine function in the $p$ regime.
In fact, this special feature of the $\epsilon$ expansion,
the existence of the constant term in particular, 
has been used for extracting the low-energy constants
from finite volume lattice QCD 
\cite{Fukaya:2007fb, Fukaya:2007pn, Fukaya:2009fh, Fukaya:2011in}. 
However, in this work, we consider a different direction:
how to reduce such peculiarity in the $\epsilon$ regime.

The answer is given by two steps. The first step is to insert
a non-zero spatial momentum ${\bf p}$ :
\begin{eqnarray}
\hspace{-0.2in}C^{\rm 2pt}_{PP}(t,{\bf p})\equiv \int d^3x e^{-i{\bf p}\cdot{\bf x}} \langle P(x)P(0) \rangle
&=& Y \frac{1}{2 E({\bf p})\sinh (E({\bf p})T/2)}\cosh (E({\bf p}) (t-T/2))+\cdots,
\end{eqnarray}
where $E({\bf p})=|{\bf p}|$.
Note that the constant $X$ automatically vanishes, and
the leading term has the same hyperbolic cosine form
as in the $p$ expansion, except for an over-all 
coefficient $Y$.
The second step is to take the ratio of them with different momenta:
$C^{\rm 2pt}_{PP}(t,{\bf p})/C^{\rm 2pt}_{PP}(t,{\bf p}^\prime)$.
Then we can eliminate $Y$ which contains Bessel functions.

Note that the sub-leading terms, expressed by ellipses, still contain
the contribution from the zero-momentum mode.
But it is suppressed typically by $\sim 1/4\pi F^2 \sqrt{V}$.

In this way, contribution from the zero-momentum pion mode
can be eliminated. It is reasonable that 
having a non-zero momentum reduces the information of the zero-mode.

\section{Three-point functions in the $\epsilon$ regime}
\label{sec:3pt}

Next,  we consider three point operators
all put at different points : ($x\neq y\neq z$).
In the $\epsilon$ expansion, the 
pseudoscalar-vector(in $0$-direction)-pseudoscalar function is given by
\begin{eqnarray}
\label{eq:3pt}
\langle P(x)V_0(y)P(z)\rangle  &=&
A \;\frac{1}{V}\sum_{p \neq 0} \frac{ip_0}{p^2} 
\left(e^{ip(x-y)}+ e^{ip(y-z)}\right)
\nonumber\\&&
+B \;\frac{1}{V}\sum_{p \neq 0}\;\frac{1}{V}\sum_{p^\prime \neq 0}
\frac{(ip_0+ip_0^\prime)e^{ip(x-y)}e^{ip^\prime(y-z)}}{p^2p^{\prime 2}}
F_V((p-p^\prime)^2)
+\cdots ,
\end{eqnarray}
where $F_V$ denotes the vector form factor, and
$A,B,\cdots$ denote (dimensionful) constants, including
the non-trivial contributions from the pion zero-mode.
Note in this case, that the constant term does not exist
because a ``vector'' cannot be made from 
the zero-momentum mode alone.

In the following, we consider inserting the initial (spatial)
momentum ${\bf p}_i$ to $P(x)$, 
and final momentum ${\bf p}_f$ to $P(z)$,
splitting the discussion into three different cases.
\begin{itemize}
\item [] \underline{Case 1}: ${\bf p}_i\neq 0, {\bf p}_f\neq 0$

In this case, it is straightforward to obtain
\begin{eqnarray}
\hspace{-0.3in}C^{\rm 3pt}_{PVP}(t,t^\prime ; {\bf p}_i, {\bf p}_f) 
&\equiv& 
\int d^3x \;e^{-i{\bf p}_i\cdot{\bf x}} \int d^3z \;e^{i{\bf p}_f\cdot{\bf z}}
\langle P(x)V_0(y)P(z)\rangle
= B F_V(q^2) 
\nonumber\\&&\hspace{-1in}\times
[E({\bf p}_i)+E({\bf p}_f)]
\left(
\frac{\cosh (E({\bf p}_i) (t-T/2))}
{2 E({\bf p}_i)\sinh (E({\bf p}_i)T/2)}
\right)
\left(
\frac{\cosh (E({\bf p}_f) (t^\prime-T/2))}
{2 E({\bf p}_f)\sinh (E({\bf p}_f)T/2)}
\right)+\cdots,
\end{eqnarray}
where $q^2=({\bf p}_i-{\bf p}_f)^2-(E({\bf p}_i)-E({\bf p}_f))^2$,
and we have assumed $t=x_0-y_0 < T/2$, and $t^\prime=y_0-z_0 < T/2$,
and used $\sinh (E({\bf p}_i) (t-T/2)) \sim - 
\cosh (E({\bf p}_i) (t-T/2))+{\cal O}(e^{E({\bf p}_i) (t-T/2)})$,
and a similar relation for $\sinh (E({\bf p}_f) (t^\prime-T/2))$.

Therefore, the ratio of the three-point and two two-point correlators,
\begin{eqnarray}
R_V(t, t^\prime ; |{\bf p}_i|, |{\bf p}_f|, q^2)
&\equiv&
\frac{\displaystyle \frac{1}{N^{\rm 3pt}_{|{\bf p}_i|, |{\bf p}_f|}}
\sum_{{\rm fixed}|{\bf p}_i|, |{\bf p}_f|, q^2}
\frac{C^{\rm 3pt}_{PVP}( t, t^\prime ; {\bf p}_i, {\bf p}_f)}
{E({\bf p}_i)+E({\bf p}_f)}}
{\displaystyle\left(\frac{1}{N^{\rm 2pt}_{|{\bf p}_i|}}
\sum_{{\rm fixed}|{\bf p}_i|}C^{\rm 2pt}_{PP}( t,{\bf p}_i)\right)
\left(\frac{1}{N^{\rm 2pt}_{|{\bf p}_f|}}
\sum_{{\rm fixed}|{\bf p}_f|}C^{\rm 2pt}_{PP}( t^\prime,{\bf p}_f)\right)}
\nonumber\\
&=&\frac{B}{Y^2}F_V(q^2)+\cdots,
\end{eqnarray}
is a good quantity to extract the form factor.
Note that the rotationally symmetric
correlators are averaged ($N^{\rm 3pt}_{|{\bf p}_i|, |{\bf p}_f|}$ and 
$N^{\rm 2pt}_{|{\bf p}_i|}$ denote the numbers of correlators in the summations).

\item [] \underline{Case 2}: ${\bf p}_i\neq 0, {\bf p}_f = 0$ 
(or ${\bf p}_i = 0, {\bf p}_f \neq 0$)

Let us next consider the case with ${\bf p}_f =0$, 
but the initial momentum is kept non-zero.
(the opposite case with ${\bf p}_i = 0, {\bf p}_f \neq 0$, 
is obtained by simply replacing ${\bf p}_f = - {\bf p}_i$).

In this case, the contribution from the first term 
of Eq.~(\ref{eq:3pt}) remains.
It is, however, easy to cancel it by taking a subtraction:
\begin{eqnarray}
\hspace{-0.3in}
\Delta_{t^\prime} C^{\rm 3pt}_{PVP}(t,t^\prime ; {\bf p}_i, 0) &\equiv& 
C^{\rm 3pt}_{PVP}(t,t^\prime ; {\bf p}_i, 0)
-C^{\rm 3pt}_{PVP}(t,t_{\rm ref} ; {\bf p}_i, 0)
\nonumber\\
&&\hspace{-0.5in}=
B F_V(q^2) 
\left(
\frac{\cosh (E({\bf p}_i) (t-T/2))}
{2 E({\bf p}_i)\sinh (E({\bf p}_i)T/2)}
\right)
\nonumber\\&&
\hspace{-0.5in}
\times \left[
-\left(
h^\prime_1 \left(\frac{t^\prime}{T}\right)
-h^\prime_1 \left(\frac{t_{\rm ref}}{T}\right)
\right)
+E({\bf p}_i)T 
\left(h_1\left(\frac{t^\prime}{T}\right)
-h_1\left(\frac{t_{\rm ref}}{T}\right)\right)\right]
+\cdots ,
\end{eqnarray}
where 
\begin{eqnarray}
h_1(\tau)\equiv \frac{1}{2}\left(\tau-\frac{T}{2}\right)^2-\frac{1}{24},
\end{eqnarray}
and $h_1^\prime (\tau)$ is its $\tau$-derivative.
The reference time $t_{\rm ref}$ can be taken arbitrarily.

For having a long-range correlation due to the zero-mode,
and a periodic temporal extent,
the correlator may have contamination from
the $\langle {\rm vac} |V_0|\pi \pi \rangle$ matrix element
wrapping around the lattice.
In fact, its contribution is not negligible 
at the leading order of ChPT.
But including the next-to-leading order corrections,
it is suppressed by a factor $\sim e^{-2 m_\pi |t^\prime -T/2|}$,
and, therefore, neglected here.

Using $C^{\rm 2pt}_{PP}( t ,0)\propto T h_1(t/T)$
and its $t$-derivative defined by 
$\partial C^{\rm 2pt}_{PP}( t,0) \propto h^\prime_1(t/T)$
in the $\epsilon$ regime, let us define
\begin{eqnarray}
\hspace{-0.3in}R_V^1(t, t^\prime ; |{\bf p}_i|, 0, q^2)
&\equiv&
\frac{\displaystyle \frac{1}{N^{\rm 3pt}_{|{\bf p}_i|}}
\sum_{{\rm fixed}|{\bf p}_i|, q^2}
\left(C^{\rm 3pt}_{PVP}( t, t^\prime ; {\bf p}_i,0)
-C^{\rm 3pt}_{PVP}( t, t_{\rm ref} ; {\bf p}_i,0)\right)}
{\displaystyle\frac{1}{N^{\rm 2pt}_{|{\bf p}_i|}}
\sum_{{\rm fixed}|{\bf p}_i|}C^{\rm 2pt}_{PP}( t,{\bf p}_i)
\left[
-\Delta_{t^\prime} \partial C^{{\rm 2pt}}_{PP}( t^\prime ,0)
+E({\bf p}_i)\Delta_{t^\prime} C^{{\rm 2pt}}_{PP}( t^\prime ,0)
\right]},
\end{eqnarray}
where 
$\Delta_{t^\prime} \partial C^{{\rm 2pt}}_{PP}( t^\prime ,0) 
\equiv \partial C^{{\rm 2pt}}_{PP}( t^\prime ,0)- 
\partial C^{{\rm 2pt}}_{PP}( t_{\rm ref} ,0)$, 
and $\Delta_{t^\prime} C^{{\rm 2pt}}_{PP}( t^\prime ,0) 
\equiv C^{{\rm 2pt}}_{PP}( t^\prime ,0)- C^{{\rm 2pt}}_{PP}( t_{\rm ref} ,0)$,
which is again expected to be $B F_V(q^2)/Y^2$ at large separations of $t$, and $t^\prime$.
Here, $q^2=p_i^2$.

\item [] \underline{Case 3}: ${\bf p}_i = {\bf p}_f = 0$

Now it is not difficult to see
\begin{eqnarray}
R_V^{2}(t, t^\prime ; 0, 0, 0)
&\equiv&
\frac{\displaystyle 
\Delta_t \Delta_{t^\prime} C^{\rm 3pt}_{PVP}( t, t^\prime ; 0,0)}
{\displaystyle
-\Delta_t C^{\rm 2pt}_{PP}(t,0)\Delta_{t^\prime} 
\partial C^{{\rm 2pt}}_{PP}( t^\prime ,0)
-\Delta_t \partial C^{\rm 2pt}_{PP}(t,0)\Delta_{t^\prime} C^{{\rm 2pt}}_{PP}( t^\prime ,0)},
\end{eqnarray}
where we have defined
$
\Delta_t \Delta_{t^\prime} C^{\rm 3pt}_{PVP}( t, t^\prime ; 0,0)\equiv C^{\rm 3pt}_{PVP}( t, t^\prime ; 0,0)
-C^{\rm 3pt}_{PVP}( t, t_{\rm ref} ; 0,0)
$

$
-C^{\rm 3pt}_{PVP}( t_{\rm ref}, t^\prime ; 0,0)
+C^{\rm 3pt}_{PVP}( t_{\rm ref}, t_{\rm ref} ; 0,0),
$ is a good quantity to extract $BF_V(0)/Y^2$.
Fortunately, there is no contamination from 
the $\langle {\rm vac} |V_0|\pi \pi \rangle$ matrix element
in this correlator.
\end{itemize}


Since $R_V$, $R_V^1$, and $R_V^{2}$ share 
the same overall factor, $B/Y^2$, one can eliminate
it by taking their ratios.
Noting $F_V(0)=1$, we find the following two independent extractions
\begin{eqnarray}
F_V(t,t^\prime,q^2)&\equiv & 
\frac{R_V(t, t^\prime ; |{\bf p}_i|, |{\bf p}_f|, q^2)}
{R_V^{2}(t, t^\prime ; 0, 0, 0)},\;\;\;
F^1_V(t,t^\prime, q^2)\equiv  
\frac{R^1_V(t, t^\prime ; |{\bf p}_i|, 0, q^2)}
{R_V^{2}(t, t^\prime ; 0, 0, 0)},
\end{eqnarray}
with $t_{\rm ref}=T/4=12$ (on our lattice),
are numerically good choices, in that 
the signals look clean, while the contamination from
the excited states is expected to be small.

\section{Preliminary lattice results}
\label{sec:lattice}

Our numerical simulations are performed with
the Iwasaki gauge action at $\beta=2.3$ and 2+1 dynamical flavors of
overlap quark action on a $16^3\times 48$ lattice.
The lattice cutoff $1/a$ = 1.759(8)(5)~GeV ($a\sim$ 0.112(1) fm) is 
determined from the $\Omega$-baryon mass.
The physical lattice size is $L\sim 1.8$ fm.

In this work, we focus on an ensemble with
the smallest up-down quark mass, $ma=0.002$.
This value roughly corresponds to 3~MeV 
and the pion mass is  $m_\pi \sim 99$ MeV \cite{Fukaya:2011in}
which is below the physical point. 
For the strange quark, we choose its mass almost 
at the physical value, $m_s a$ = 0.080.
Note in this set up, the pions are in the $\epsilon$ regime
($m_\pi L\sim 0.90$).

In the Hybrid Monte Carlo (HMC) updates, the global topological
charge of the gauge field is fixed to $Q=0$.
Since its effect is encoded in the pion zero-mode,
the $Q$ dependence should not appear in the ratios of our correlators.

For the computation of the correlation functions,
we use the smeared sources with the form of a single exponential function.
To improve the statistical signal, the so-called all-to-all propagator
technique is used:
the low energy part of the correlator 
is separately calculated by the 160 eigenmodes of the Dirac operator 
and averaged over different source points, and
the higher-mode contribution is estimated stochastically 
by the noise method with the dilution technique.

Details of the numerical simulation will be reported elsewhere.

\begin{figure*}[tb]
  \centering
  \includegraphics[width=7.5cm]{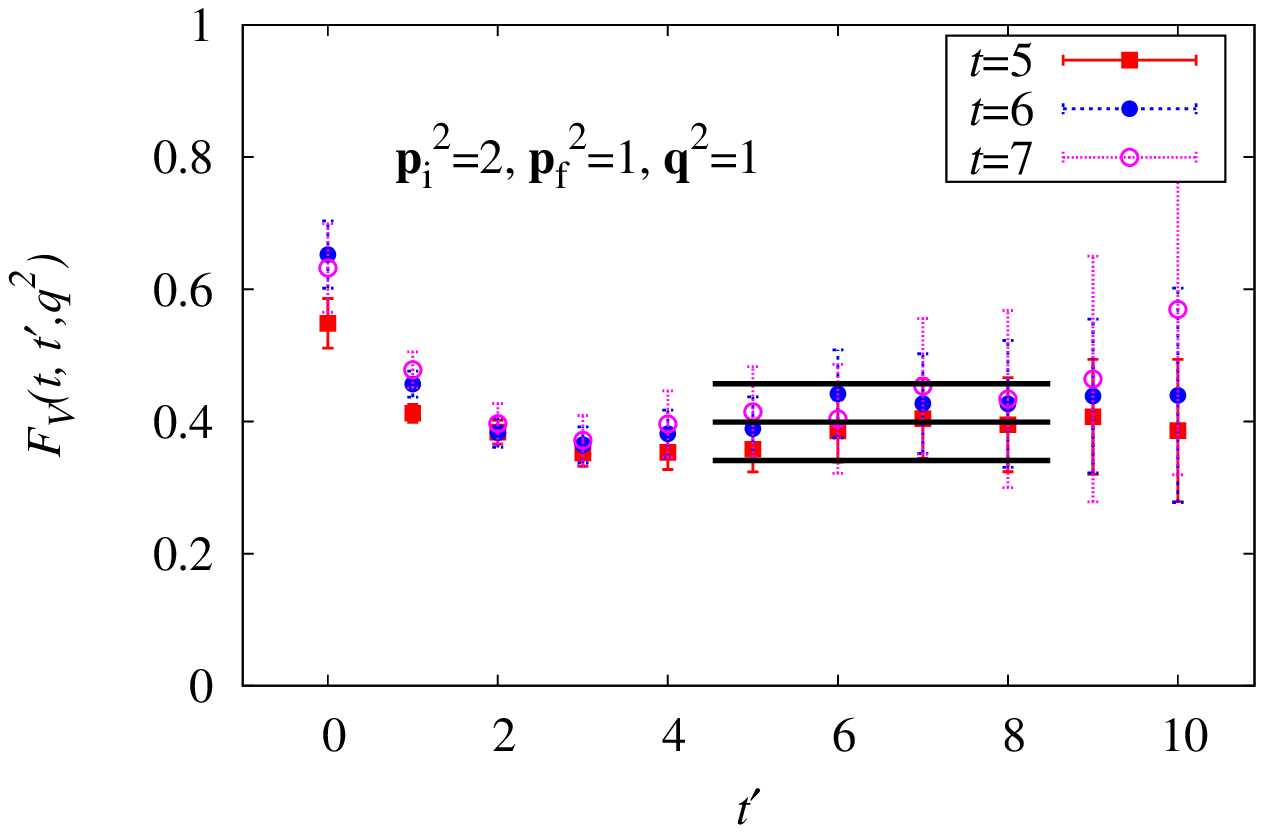}
  \includegraphics[width=7.5cm]{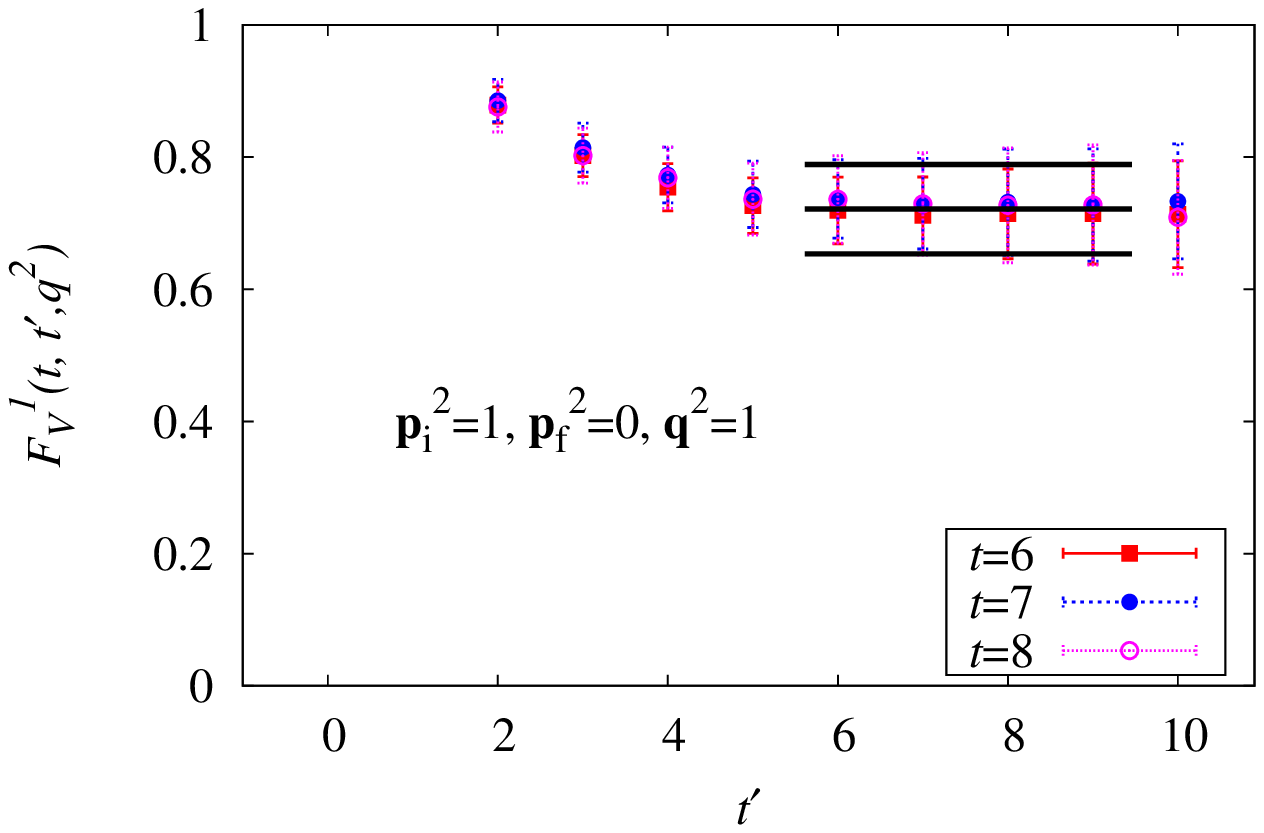}
  \caption{$F_V(t,t^\prime,q^2=-0.11[\mbox{GeV}^2])$ 
(left panel) and $F_V^1(t,t^\prime,q^2=-0.40[\mbox{GeV}^2])$ (right).}
  \label{fig:FV}
\end{figure*}

Figure~\ref{fig:FV} shows our lattice data
of $F_V(t,t^\prime,q^2)$ at $({\bf p}_i^2, {\bf p}_f^2, {\bf q}^2)=(2,1,1)$ 
(left panel) and $F_V^1(t,t^\prime,q^2)$ at 
$({\bf p}_i^2, {\bf p}_f^2, {\bf q}^2)=(1,0,1)$
(right panel). Here, the momenta are denoted in units of $2\pi /L$.
To estimate $E({\bf p})$, we have used the dispersion relation
$E({\bf p})=\sqrt{{\bf p}^2+m_\pi^2}$.
Each set of data shows a plateau around $t,t^\prime = 5$--9 
from which we extract $F_V(q^2)$ by a constant fit.

\begin{wrapfigure}[17]{r}{0.57\linewidth}
  \centering
  \vspace{-0.4in}
  \includegraphics[width=8cm]{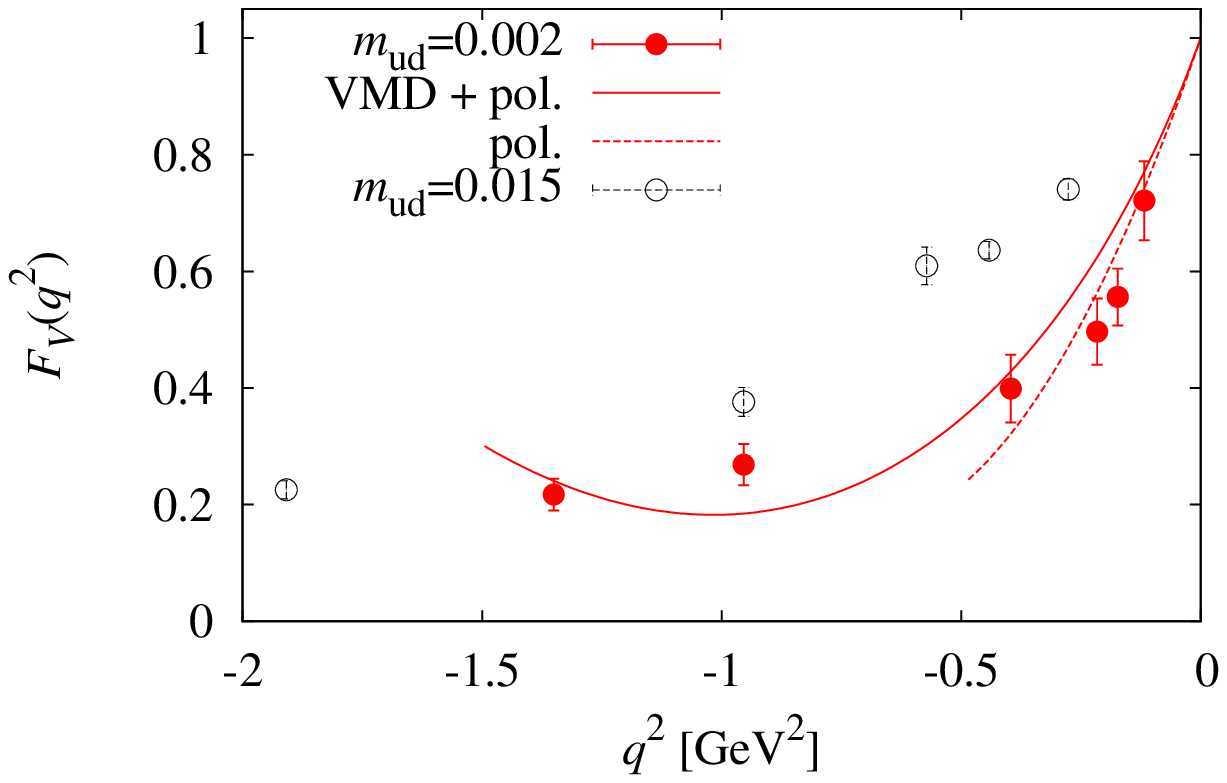}
  \caption{The $q^2$ dependence of $F_V(q^2)$.}
  \label{fig:result}
  \vspace{0.25in}
  \includegraphics[width=7.5cm]{r2.eps}
  \caption{The $m_\pi^2$ dependence of the pion charge radius.}
  \label{fig:result2}
\end{wrapfigure}

The $q^2$ dependence of $F_V(q^2)$ 
is presented in Fig.~\ref{fig:result}.
For comparison, we have also plotted 
data in the $p$ regime (at $m = 0.015$) \cite{Kaneko:2010ru}.
Our new data in the $\epsilon$ regime shows
a steeper slope near the origin,
which results in a large value of the pion charge radius,
\begin{equation}
\langle r^2\rangle_V\equiv 6 \frac{\partial F_V(q^2)}{\partial q^2}
= 0.53(4)\;\;\mbox{fm}^2 \;\;[\mbox{preliminary}].
\end{equation}
This value is obtained by fitting the data
to a function 
$
F_V(q^2)=\frac{1}{1-q^2/m_\rho^2} + a_1 q^2 +a_2 (q^2)^2
$
\cite{Aoki:2009qn}, (solid curve in Fig.~\ref{fig:result})
with an input of the rho-meson mass $m_\rho =770$ MeV
(a fit without rho-meson contribution (dashed curve) is also tried).
This result is larger than the experimental value ($0.452(11)$fm$^2$)
and confirms the existence of the strong (logarithmic)
curvature of the pion charge radius near the chiral limit,
as shown in Fig.~\ref{fig:result2}.\\

In this work, we have demonstrated 
how to cancel finite volume effects in the $\epsilon$ regime.
Inserting momenta to the operators,
and taking appropriate ratios of them,
we can eliminate the contribution from the
pion zero-mode.
This cancellation occurs
only at the leading order, and
there should be non-trivial corrections from NLO terms.
But it is suppressed by a factor $\sim 1/4\pi F^2 \sqrt{V}$
($\sim 7$\% on our lattice).
Our tree-level analysis of the vector pion form factor
on the lattice in the $\epsilon$ regime
confirms the above observation and our preliminary 
result for the pion charge radius 
is consistent with the experiment,
showing the existence of a diverging logarithmic curve.\par

In this work, we have focused on the calculation of the 
vector (or electro-magnetic) form factor.
But in principle, this method can be applied 
for any other form factors.\\

We thank P.~H.~Damgaard for useful discussions.
Numerical simulations are performed on the IBM System Blue Gene
Solution at High Energy Accelerator Research Organization
(KEK) under a support of its Large Scale Simulation Program (No. 09/10-09, 11-05).
This work is supported in part by the Grant-in-Aid of the
  Japanese Ministry of Education
  (No. 21674002, 22740183), 
the Grant-in-Aid for Scientific
Research on Innovative Areas (No. 2004: 20105001, 20105002, 20105003, 20105005, 23105710),
and SPIRE (Strategic Program for Innovative Research).

\end{document}